\documentclass[superscriptaddress,twocolumn,showpacs,amsmath,amssymb]{revtex4}

\usepackage{graphicx}
\usepackage{dcolumn}
\usepackage{bm}

\begin{document}

\preprint{}

\title{Probing surface diffuseness of nucleus-nucleus potential\\
with quasielastic scattering at deep sub-barrier energies}

\author{K. Washiyama}
\affiliation{
Department of Physics, Tohoku University,
Sendai 980-8578, Japan}
\author{K. Hagino}
\affiliation{
Department of Physics, Tohoku University,
Sendai 980-8578, Japan}
\author{M. Dasgupta}
\affiliation{
Department of Nuclear Physics, Research School of Physical Sciences 
and Engineering, Australian National University, Canberra, ACT 0200, 
Australia}

\date{\today}

\begin{abstract}
We perform a systematic study
on the surface property of nucleus-nucleus potential in heavy-ion 
reactions using 
large-angle quasielastic scattering
at energies well below the Coulomb barrier. 
At these energies, the quasielastic scattering can be well described 
by a single-channel potential model. 
Exploiting this fact, 
we point out that 
systems which involve spherical nuclei
require the diffuseness parameter 
of around 0.60 fm
in order to fit the experimental data, 
while systems with a deformed target
between 0.8 fm and 1.1 fm.
\end{abstract}

\maketitle

\section{Introduction}
The Woods-Saxon form,
which is characterized by the 
depth, radius and diffuseness parameters,
has often been used for the 
inter-nuclear potential for heavy-ion reactions.
Conventionally, 
the diffuseness parameter 
of around 0.63 fm has been employed 
for calculations of 
elastic and inelastic scattering,
which are sensitive only to the surface region of 
the nuclear potential \cite{broglia91,Christensen76}.
This value of surface diffuseness parameter has been well accepted, 
partly because it is consistent with 
a double folding potential \cite{SL79}. 
In contrast, 
a recent systematic study has shown that
experimental data for heavy-ion fusion reactions at energies
close to the Coulomb barrier require 
a larger value of the diffuseness parameter, 
ranging between 0.75 and 1.5 fm, 
as long as the Woods-Saxon 
parameterization is used as a nuclear potential \cite{newton04}.
The origin of the discrepancy in the surface diffuseness parameter 
between the scattering and fusion processes 
has not yet been understood. 

Large-angle quasielastic scattering at deep sub-barrier energies
provides an alternative way to look at this problem. 
Quasielastic scattering and fusion are both inclusive processes and 
are complimentary to each other. The former is related to the 
reflection probability at the Coulomb barrier, while the latter 
to the penetration probability. 
In heavy-ion reactions at energies near the Coulomb barrier,
it is well known that 
the channel coupling effects caused by the 
collective inelastic excitations of the colliding nuclei
strongly affect the reaction dynamics \cite{nanda98,BT98}. 
At deep sub-barrier energies, however, 
the channel coupling effects on 
quasielastic scattering can be disregarded, since the reflection 
probability is almost unity at these energies irrespective of 
the presence of channel couplings, even though inelastic 
channels themselves may be strongly populated \cite{hagino05}. 
This is similar to fusion at energies well above the Coulomb barrier, 
where the penetrability is almost unity \cite{newton04}.

The above concept was recently applied
to the experimentally measured quasielastic scattering cross sections
for the $^{16}$O + $^{154}$Sm system 
at deep sub-barrier energies \cite{hagino05}. 
It was found that the larger 
surface diffuseness parameter of around 1.0 fm 
is required for this system 
in order to fit the data. 
This value is consistent with the one required for 
fusion. 

It is apparent that a more systematic study is necessary, 
in order to clarify whether 
the quasielastic scattering 
around the Coulomb barrier generally requires a larger value of surface
diffuseness 
parameter than the conventional value of around 0.63 fm. 
The aim of this paper is 
to carry out such systematic study on quasielastic scattering at 
deep sub-barrier energies. 
To this end,
we calculate the excitation function of 
the quasielastic cross sections 
for systems involving both spherical and deformed nuclei. 
The reactions $^{32,34}$S+$^{197}$Au, $^{32,34}$S+$^{208}$Pb, 
$^{16}$O+$^{154}$Sm, $^{186}$W, $^{208}$Pb,
for which
experimental data exist at deep sub-barrier energies, are
studied. We show that a surface diffuseness parameter of around 0.6 fm
is favored by the data for reactions involving spherical nuclei,
whilst those involving deformed nuclei require a larger value of
the diffuseness parameter.

The paper is organized as follows.
In the next section, 
we briefly review the large-angle quasielastic scattering 
at deep sub-barrier energies. 
We also explain the procedure of our analyses which use a 
one dimensional ion-ion potential, including our definition
of deep sub-barrier energies. 
In Sec. III, we present our results for the $\chi^2$ fitting 
and discuss its sensitivity to the barrier height energy 
and to the channel coupling effects.
We summarize the paper in Sec. IV.

\section{Method of analyses}
\subsection{Large-angle quasielastic scattering at deep sub-barrier energies}

Our purpose in this paper 
is to study the surface property of ion-ion potential 
using 
heavy-ion quasielastic scattering.  
Before we explain the method of our analyses, 
let us first discuss briefly 
the advantage of exploiting large-angle quasielastic scattering 
at deep sub-barrier energies. 

At energies well below the Coulomb barrier, 
the cross sections of (quasi)elastic scattering are close to the 
Rutherford cross sections, with small deviations 
caused by the effect of nuclear interaction. 
This effect can be taken into account 
by the semiclassical perturbation theory. 
The ratio of elastic scattering $\sigma_{\rm el}$
to Rutherford cross sections $\sigma_R$ at a backward angle $\theta$
is given by \cite{hagino04,LW81}
\begin{equation}
\frac{d\sigma_{\rm el}(E_{cm},\theta)}{d\sigma_R(E_{cm},\theta)}
\sim
1+\frac{V_N(r_c)}{ka}\,
\frac{\sqrt{2a\pi k\eta}}{E_{cm}}, 
\label{deviation}
\end{equation}
where 
$E_{cm}$ is the centre-of-mass energy,
$k=\sqrt{2\mu E_{cm}}/\hbar$, $\mu$ being the reduced mass, 
and $\eta$ is the Sommerfeld parameter. 
This formula is obtained by assuming 
that the nuclear potential $V_N(r)$ has an exponential form,
$\exp(-r/a)$,
around the classical turning point 
$r_c=(\eta+\sqrt{\eta^2+\lambda_c^2})/k$,
where 
$\lambda_c=\eta \cot(\theta/2)$ is the classical angular momentum 
for the Rutherford scattering. 
We see from this formula
that the deviation of the elastic cross sections
from the Rutherford ones is sensitive to the surface region 
of the nuclear potential, especially to the surface diffuseness 
parameter $a$. 
Notice that, for 
small scattering angles, the Fresnel oscillation
may complicate the formula. 
Also, as mentioned in the previous section, 
the channel coupling effects on the quasielastic cross sections 
are negligible at deep sub-barrier energies. 
We can thus study the effect of the surface diffuseness parameter
in a transparent and unambiguous way using the large-angle quasielastic
scattering at deep sub-barrier energies. 

\subsection{Procedure}

In order to compare with the experimental data for 
the quasielastic cross sections
at deep sub-barrier energies,
we use a one-dimensional optical potential 
with the Woods-Saxon form. 
Absorption following transmission through the barrier
is simulated by an imaginary potential with
$W=30$ MeV, $a_w=0.4$ fm, and $r_w=1.0$ fm.
This model calculates the elastic and 
fusion cross sections, in which the elastic cross sections can be 
considered as quasielastic cross sections to a good approximation 
at these deep sub-barrier energies \cite{comment}. 
Note that the results are insensitive to the parameters of 
the imaginary part 
as long as it is well localized inside the Coulomb barrier.

In order to carry out a systematic study,
we calculate the Coulomb barrier height using 
the Aky\"uz-Winther potential \cite{akyuz81}. 
We examine several potentials with different values of surface 
diffuseness parameter, which give the same calculated barrier height. 
To this end, 
we vary the radius parameter $r_0$ 
while keeping the depth parameter $V_0$ to be 100 MeV. 
This is possible 
because the effect of variation in $V_0$ and $r_0$ 
on the Coulomb barrier height compensates with each other 
at the surface region. 

We define the region of ``deep sub-barrier energies''
in the following way. 
In heavy-ion collisions at energies near the Coulomb barrier,
collective inelastic excitations of the colliding nuclei 
and transfer reactions 
are strongly coupled to the relative motion.
This causes 
the splitting of the Coulomb barrier into several distributed 
barriers \cite{rowley91,nanda98}. 
We define the deep sub-barrier energies as 
around 3 MeV below the lowest barrier height or smaller. 
For this purpose, we first use the computer code
CCFULL \cite{HRK99} in order to 
explicitly construct the coupling matrix 
(which includes the excitation energy for the diagonal components) 
for the coupled-channels equations 
for each system by including known low-lying 
collective excitations.  
We then diagonalize it to obtain the lowest eigen-barrier. 

We find that the deep sub-barrier region defined in this way 
corresponds to 
the region where
the experimental value of the ratio of the quasielastic 
to the Rutherford cross sections
is larger than around 0.94.
We therefore include 
only those experimental data which satisfy 
$d\sigma_{\rm qel}/d\sigma_R\ge 0.94$ 
in the $\chi^2$ fitting. 
A few experimental data points with values exceeding unity 
were excluded while performing the fits, 
but are shown in the figures below.

We apply this procedure to the 
$^{32,34}$S + $^{208}$Pb, 
$^{32,34}$S + $^{197}$Au \cite{schuck02}, 
and $^{16}$O + $^{208}$Pb \cite{timmers96} 
reactions which involve spherical nuclei, as 
well as the 
$^{16}$O + $^{154}$Sm and $^{16}$O + $^{186}$W reactions
\cite{timmers95} which involve a deformed target. 
For the
deformed systems the scarcity of data points at deep sub-barrier
energies led us to extend the fitting region to somewhat higher
energies. This meant that the calculations had to take account of
deformation effects as explained in Sec. III. B.

\section{Results and discussion}

\subsection{Spherical systems}

\begin{figure}[tbhp]
\begin{center}\leavevmode
\includegraphics[width=0.95\linewidth, clip]{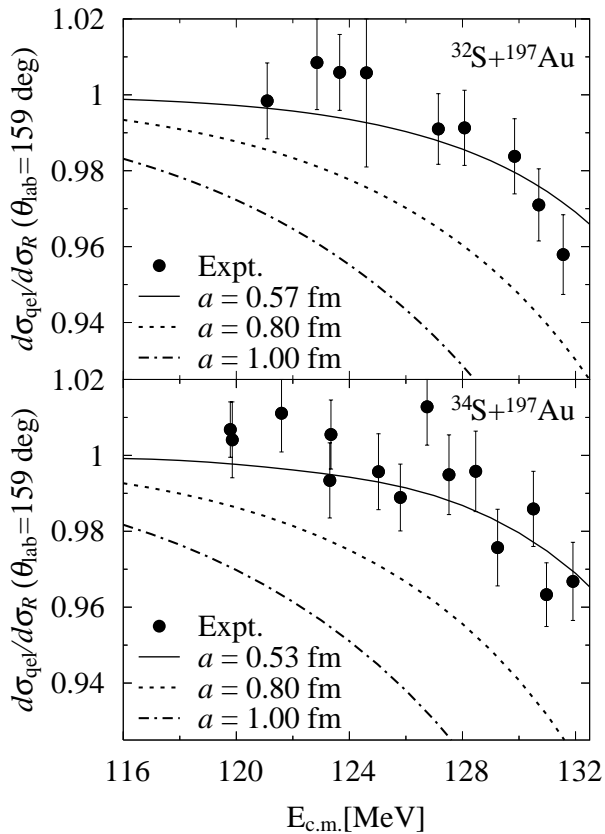}
\caption{
The ratio of the quasielastic to the Rutherford cross sections
at $\theta_{\rm lab}=159^\circ$  
for the $^{32}$S + $^{197}$Au (the upper panel) reaction
and for the $^{34}$S + $^{197}$Au (the lower panel) reaction.
The experimental data are taken from Ref. \cite{schuck02}. 
The solid line results from using a diffuseness parameter obtained by 
performing a least-square fit to the data.
The dotted and the dot-dashed lines are obtained with 
the diffuseness parameter of $a$ = 0.80 fm and $a$ = 1.00 fm,
respectively. 
}
\label{fig:SandAu}
\end{center}
\end{figure}

We first present the results 
for systems involving spherical nuclei.
Figure 1 compares the experimental data with
the calculated cross sections obtained with 
different values of the surface diffuseness 
parameter in the Woods-Saxon potential 
for the $^{32}$S + $^{197}$Au system (the upper panel) 
and the $^{34}$S + $^{197}$Au system (the lower panel). 
The Coulomb barrier height 
is 141.2 MeV for the $^{32}$S + $^{197}$Au reaction
and is 140.2 MeV for the $^{34}$S + $^{197}$Au reaction. 
The best fitted values for the surface diffuseness parameter 
are $a=0.57\pm 0.04$ fm and $a=0.53\pm 0.03$ fm 
for the $^{32}$S and $^{34}$S + $^{197}$Au reactions, respectively. 
The cross sections obtained with these surface diffuseness 
parameters are denoted by the solid line in the figure. 
The dotted and the dot-dashed lines are calculated with
the diffuseness parameter of $a$ = 0.80 fm and $a$ = 1.00 fm,
respectively.
Figure 2 shows the results for 
the $^{32}$S + $^{208}$Pb (the upper panel) 
and the $^{34}$S + $^{208}$Pb (the lower panel) reactions. 
The Coulomb barrier height is 145.1 MeV and 
144.1 MeV 
for the $^{32}$S and $^{34}$S 
+ $^{208}$Pb reactions, respectively. 
The best fitted values for the surface diffuseness 
parameter are 
$a=0.60\pm 0.04$ fm and $a=0.63\pm 0.04$ fm 
for the $^{32}$S and $^{34}$S 
+ $^{208}$Pb reactions, respectively. 

\begin{figure}[tbhp]
\begin{center}\leavevmode
\includegraphics[width=0.95\linewidth, clip]{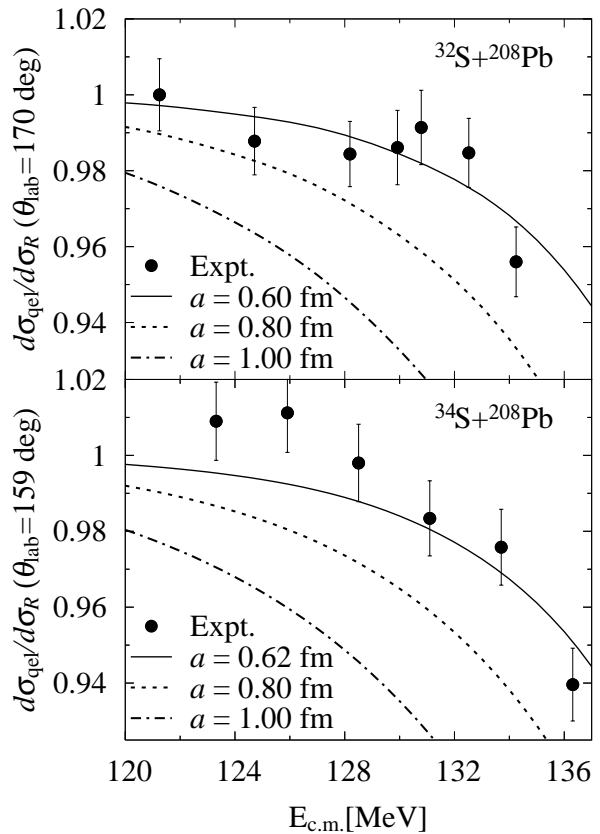}
\caption{
The ratio of the quasielastic to the Rutherford cross sections
for the $^{32}$S + $^{208}$Pb (the upper panel) reaction
at $\theta_{\rm lab}=170^\circ$  
and for the $^{34}$S + $^{208}$Pb (the lower panel) reaction
at $\theta_{\rm lab}=159^\circ$.
The experimental data are taken from Ref. \cite{schuck02}. 
The meaning of each line is the same as in Fig. 1.
}
 \label{fig:s-pb}
\end{center}
\end{figure}

\begin{figure}[tbhp]
\begin{center}\leavevmode
\includegraphics[width=0.95\linewidth, clip]{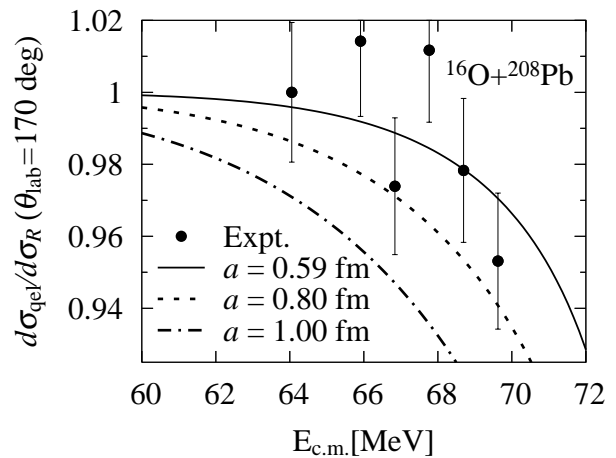}
\caption{
The ratio of the quasielastic to the Rutherford cross sections
for the $^{16}$O + $^{208}$Pb reaction
at $\theta_{\rm lab}=170^\circ$. 
The experimental data are taken from Ref. \cite{timmers96}. 
The meaning of each line is the same as in Fig. 1.
}
 \label{fig:o-pb}
\end{center}
\end{figure}

It is evident from Figs. 1 and 2 that 
these spherical systems 
favor the standard value of the surface diffuseness parameter,
around $a=$ 0.60 fm. 
The calculations with the larger diffuseness 
parameters, $a$ = 0.80 fm and 1.00 fm, 
underestimate the quasielastic cross sections 
and are not consistent with the energy dependence of the 
experimental data.
We obtain a similar conclusion for the 
$^{16}$O + $^{208}$Pb system, where the best fitted value for 
the surface diffuseness parameter is $a=0.59 \pm 0.10$ fm
with the Coulomb barrier height of 76.1 MeV. 
The result for this system is shown in  Fig. \ref{fig:o-pb}.

\begin{figure}[tbhp]
\begin{center}\leavevmode
\includegraphics[width=0.95\linewidth, clip]{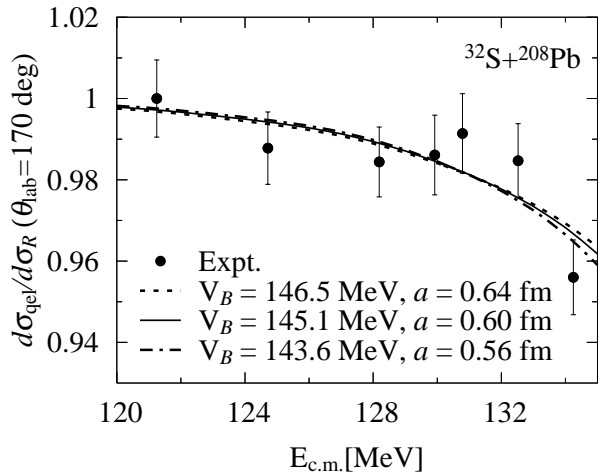}
\caption{
Comparison of 
quasielastic cross sections 
obtained for three different values of the Coulomb barrier height 
for the $^{32}$S + $^{208}$Pb reaction. 
The surface diffuseness parameter is determined for each barrier 
energy by fitting the experimental data. 
}
\label{fig:SandPbcompare}
\end{center}
\end{figure}

The 
conclusions are not sensitively dependent on the choice of 
barrier height energy $V_B$. 
In order to demonstrate this, we vary the barrier height 
by 1\%, and repeat the same analyses. 
The result for the $^{32}$S + $^{208}$Pb system 
is shown in Fig. \ref{fig:SandPbcompare}.
The solid line denotes the result obtained 
with the Aky\"uz-Winther potential, as a reference, which is the same 
as the solid line in the upper panel of Fig. 2. 
The best fits and the resulting $a$ values using $V_B$ = 143.6 MeV and 
$V_B$ = 146.5 MeV are also shown in Fig. \ref{fig:SandPbcompare}. 
The $a$ value changes by $\pm0.04$ fm 
for a $\pm 1\%$ change in the barrier energy. 
The cross sections obtained with these potentials are 
shown in the figure by the dotted and the dot-dashed lines,
respectively. 
One clearly sees that the effect of the variation of the 
Coulomb barrier height on 
the surface diffuseness parameter is small. 
The barrier energy obtained from the analysis of 
the above-barrier fusion cross sections is 144.03 MeV \cite{newton04},
which is within the range of $V_B$ used in the calculations. Thus, the 
diffuseness parameter extracted in this work will not change 
significantly if $V_B$ determined from fusion data, instead of the 
Aky\"uz-Winther prescription, is used.
We have confirmed a similar behavior of the surface diffuseness 
parameter $a$  for the other systems 
as well. 

\subsection{Deformed systems}

Let us next discuss the systems with a deformed target, 
that is, $^{16}$O + $^{154}$Sm, $^{186}$W reactions. 
For these systems, only a few data points are available at
deep sub-barrier energies.
We therefore include 
the experimental data at energies not only well below but also 
around the lowest barrier  
in the $\chi^2$ fitting procedure. 
At these energies, 
the channel coupling effects start playing an important role in 
quasielastic reactions, 
and we 
include the effect of deformation of the target nucleus 
in our calculations. 
Therefore, our analyses for the deformed systems are 
somewhat more model dependent than 
those for the spherical systems 
presented in the previous subsection. 

In order to account for the deformation effect on the 
quasielastic scattering, 
we use the orientation average formula \cite{hagino04,ARN88}, 
in which we neglect the finite excitation energy
of the ground state rotational band.
With this formula, the quasielastic cross section is given by,
\begin{equation}
\sigma_{\rm qel}(E_{cm},\theta)=\int^1_0d(\cos\theta_T)
\,\sigma_{\rm el}(E_{cm},\theta;\theta_T),
\end{equation}
where $\theta_T$ is the angle between the symmetry axis 
of the deformed target
and the direction of the projectile from the target.
In the calculation for 
both the systems,
we take six different orientation angles into account
\cite{nagara86}. 
The results change only marginally even if we include 
the larger number of orientation angles. 

\begin{figure}[tbp]
\begin{center}\leavevmode
\includegraphics[width=0.95\linewidth, clip]{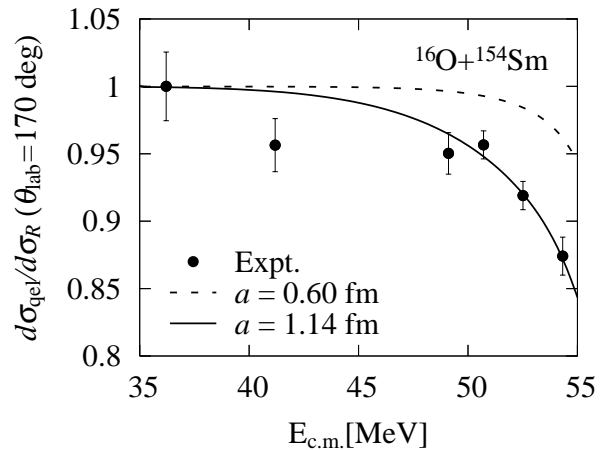}
\caption{
The ratio of the quasielastic to the Rutherford cross sections
for the $^{16}$O + $^{154}$Sm reaction
at $\theta_{\rm lab}=170^\circ$.
The solid line is obtained using 
the best fitted value of the surface diffuseness parameter, 
$a=1.14$
fm. 
The dotted line denotes the cross sections obtained with 
the diffuseness parameter of $a$ = 0.60 fm. 
The experimental data are taken from Ref. \cite{timmers95}. 
}
 \label{fig:osm}
\end{center}
\end{figure}

\begin{figure}[tbp]
\begin{center}\leavevmode
\includegraphics[width=0.95\linewidth, clip]{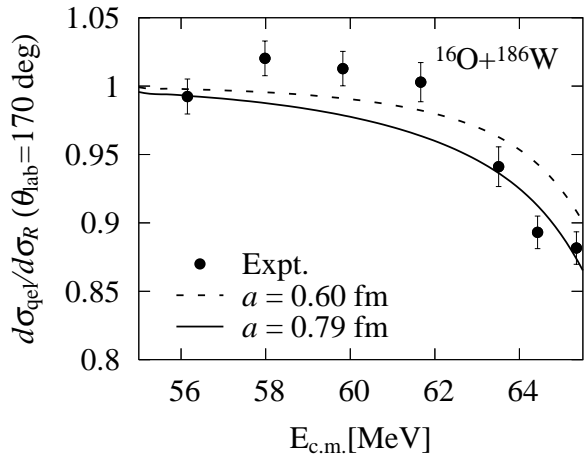}
\caption{
The ratio of the quasielastic to the Rutherford cross sections
for the $^{16}$O + $^{186}$W reaction
at $\theta_{\rm lab}=170^\circ$.
The meaning of each line is the same as in
Fig. \ref{fig:osm}.
The experimental data are taken from Ref. \cite{timmers95}. 
}
 \label{fig:ow}
\end{center}
\end{figure}

The best fitted value for the surface diffuseness 
parameter obtained in this way is $a=1.14 \pm 0.03$ fm and 
0.79 $\pm 0.04$ fm for the $^{16}$O + $^{154}$Sm 
and $^{16}$O + $^{186}$W reactions, respectively. 
The deformation parameters which we use in the calculations 
are 
$\beta_2=0.306$ and $\beta_4=0.05$ 
for $^{154}$Sm
and $\beta_2=0.29$ and $\beta_4=-0.03$ 
for $^{186}$W.
Figs. \ref{fig:osm} and \ref{fig:ow} compare the 
calculated cross sections 
with the experimental data.
The solid line in each figure 
is obtained using the best fitted value of
the diffuseness parameter. 
The dotted line shows the cross section obtained with 
the diffuseness parameter of $a=0.60$ fm as a reference.
We find that 
the larger values of the surface diffuseness 
parameter, $a=1.14$ fm and 0.79 fm,
in the nuclear potential 
are favored for these deformed system, 
in accordance with our previous conclusion in Ref. \cite{hagino05}. 
For the $^{16}$O + $^{154}$Sm reaction, the calculated cross sections 
with the standard value of the surface diffuseness parameter 
around 0.60 fm are clearly in disagreement with the experimental data.

\subsection{Discussion}

\begin{figure}[tbhp]
\begin{center}\leavevmode
\includegraphics[width=0.95\linewidth, clip]{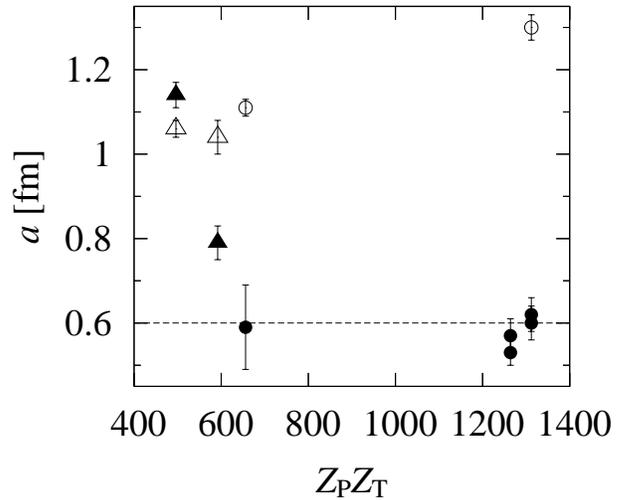}
\caption{
The best fitted values of the surface diffuseness parameter 
$a$ as a function of the charge product
of the projectile and target nuclei, $Z_{\rm P}Z_{\rm T}$.
The filled circles and triangles are for the spherical and 
the deformed systems, respectively.
The open circles and triangles are the surface diffuseness 
parameters deduced from
the analyses of fusion cross sections \cite{newton04}.
}
 \label{fig:systematics}
\end{center}
\end{figure}

Figure \ref{fig:systematics} 
summarizes the results for our systematic study 
for the surface diffuseness parameter. 
It shows the best fitted value of diffuseness parameter 
as a function of the charge product 
of the projectile and target nuclei for each system. 
The results for the spherical systems are denoted by the 
filled circles, while those for the deformed systems the filled  
triangles. One clearly sees the trend that 
the best fitted value 
of the diffuseness parameter 
is around 0.60 fm for the former, while 
it is much larger than that for the latter. 
Also, one sees that the surface diffuseness is almost constant 
for the spherical systems. 

The value of the surface diffuseness parameter obtained in this 
study for the spherical systems agrees well with the 
conventionally used value $a\sim$ 0.63 fm. 
This suggests that the double folding potential is valid at least 
in the surface region and for systems which do not involve 
a deformed target. 
For these systems, the discrepancy between the values 
of the diffuseness parameter determined from fusion data 
(open circles and triangles in Fig. \ref{fig:systematics}) 
and those from quasielastic data 
must be related with the dynamics inside the Coulomb barrier 
\cite{newton04}.

For the deformed systems studied here, the diffuseness parameter 
extracted from the quasielastic scattering 
is much larger than the conventional value of $a\sim$ 0.63 fm. 
Although this value is consistent with that extracted from 
fusion, the origin of the difference between the spherical and the 
deformed systems is not clear. 
One should bear in mind, however, that 
our analyses for the deformed systems are somewhat model dependent. 
This is due to the fact that the experimental data 
in the deep sub-barrier region are sparse for the deformed systems, 
and we need to include 
the deformation effect in the calculations in order to 
reproduce the strong energy dependence of 
the quasielastic cross sections 
at energies around the lowest barrier where the data exist. 
In order to clarify the difference in the diffuseness parameter
between the spherical and the deformed systems, 
further precision measurements for large-angle quasielastic scattering 
at deep sub-barrier energies will be necessary, especially for
deformed systems.

\section{summary}

Large-angle quasielastic scattering provides a powerful tool 
not only for the analysis of the barrier distribution 
around the Coulomb barrier 
but also for the study of the surface property of the 
nuclear potential. 
This is due to the fact that channel coupling effects play a minor role
in quasielastic scattering at deep sub-barrier energies, that enables 
a relatively model independent analysis of ion-ion potential. 
Using this fact, we have systematically analyzed 
experimental data for quasielastic scattering at deep sub-barrier 
energies, with the aim of extracting the surface diffuseness 
parameter of internuclear potential. 
We obtained the diffuseness parameter 
that is consistent with the standard value of around $a$ = 0.63 fm 
for the systems involving spherical nuclei. 
In contrast, fits to 
the data for systems involving deformed nuclei require diffuseness 
parameter to be in the range of 0.8 to 1.1 fm, similar to that 
obtained from analyses of the fusion data at above-barrier energies.

The origin of the difference between the spherical and the 
deformed systems is not clear at the moment. 
In order to clarify this and confirm the systematics found
in this paper, more experimental investigations 
on large-angle quasielastic scattering at deep sub-barrier energies 
will be certainly helpful, especially for deformed targets. 

\begin{acknowledgments}
We thank discussions with the members of the Japan-Australia Cooperative
Scientific Program ``Dynamics of Nuclear Fusion: Evolution Through a 
Complex Multi-Dimensional Landscape''. 
This work was supported by the Grant-in-Aid for Scientific Research,
Contract No. 16740139 from the Japanese Ministry of Education,
Culture, Sports, Science, and Technology.
M.D. acknowledges the support of the Australian Research Council.
\end{acknowledgments}


\begin{thebibliography}{99}
\bibitem{comment}
One should bear in mind that the cross sections calculated in 
this way may not provide a good representation for the cross sections 
of the elastic scattering themselves. For the description of
elastic scattering with a potential model,
one has to introduce an optical potential whose 
imaginary part is extended to the surface region of 
the real part in order to simulate the inelastic processes.
\end{thebibliography}
\end{document}